# Brine rejection and hydrate formation upon freezing of NaCl aqueous solutions

Ifigeneia Tsironi,[a†] Daniel Schlesinger,[b†] Alexander Späh[a], Lars Eriksson[c], Mo Segad[c] and Fivos Perakis*[a]

Studying the freezing of saltwater on a molecular level is of fundamental importance for improving freeze desalination techniques. Here, we investigate the freezing process of NaCl solutions using a combination of x-ray diffraction and molecular dynamics simulations (MD) for different salt-water concentrations, ranging from seawater conditions to saturation. A linear superposition model reproduces well the brine rejection due to hexagonal ice Ih formation and allows us to quantify the fraction of ice and brine. Furthermore, upon cooling at T = 233 K we observe the formation of NaCl·2H$_2$O hydrates (hydrohalites), which coexist with ice Ih. MD simulations are utilized to model the formation of NaCl crystallites. From the simulations we estimate that the salinity of the newly produced ice is 0.5% mass percent (m/m) due to ion inclusions, which is within the salinity limits of fresh water. In addition, we show the effect of ions on the local ice structure using the tetrahedrality parameter and follow the crystalite formation by using the ion coordination parameter and cluster analysis.

## 1. Introduction

*Freeze desalination,* also known as *cryo-desalination*, *freezing-melting* and *freeze-thaw* desalination, has been suggested as an energy effective alternative to distillation processes[1]. This is due to the fact that the latent heat of freezing (334 kJ/kg) is significantly lower than that of evaporation (2257 kJ/kg)[2]. Despite the myths and misconceptions associated with this method[3] there have been several case studies that demonstrate the feasibility of this approach from small freezing units[4] to single-step desalination plants[5], as well as for treating the waste water from Reverse Osmosis (RO) plants[6]. The energy consumption of freeze desalination has been estimated to be as low as 41.4 kJ/kg (11.5 kWh per 1000 kg of ice produced)[7] and 26.64 kJ/kg (7.4 kWh per 1000 kg of ice produced)[8], which approaches the specific energy consumption of a RO plant, typically on the order of 7 kJ/kg (about 2 kWh/m$^3$)[9,10]. The challenges and obstacles to scale-up this technology have been recently discussed[11,12] and its combination with liquefied natural gas (LNG) technologies has been considered very promising[13–17]. In addition, there have also been several studies examining freeze desalination from a physicochemical perspective, investigating the depletion of ions[18] and the ice impurities and sweating[19].

When the freezing process of salt solutions takes place, the salt ions are expelled from the solid phase, due to the preferential formation of pure ice[20]. As a result, the remaining liquid phase turns into brine, which is liquid water saturated with salt[21]. The salting-out that occurs due to freezing is one of the basic mechanisms in freeze desalination. The main challenge is the ice-brine separation, i.e. the separation of salt from crystalline ice due to the formation of polycrystalline micro-domains forming upon crystallization[22]. Experiments indicate that frozen NaCl aqueous solutions can form different NaCl-water crystal phases[23,24]. In addition, theoretical investigations indicate that on a molecular level, brine rejection progresses through a disordered layer with fluctuating ion density followed by a neat ice layer[25]. Recent simulations further indicate that ions are included in the ice lattice as dopant with Cl$^-$ ions having higher probability of inclusion than Na$^+$ ions[26,27]. Furthermore, in the case of homogeneous nucleation of saltwater, the nucleation rate decreases with addition of salt[28].

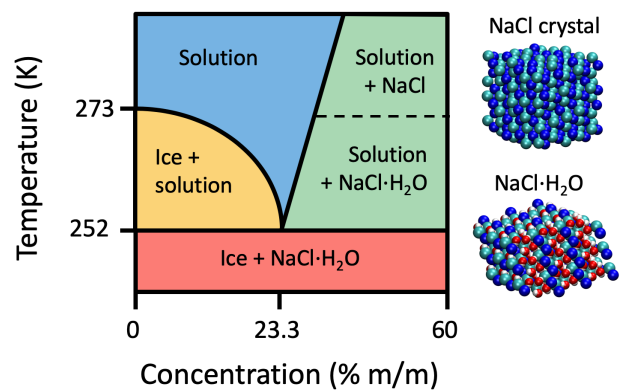

**Fig.1** A schematic phase diagram of NaCl aqueous solution. Above the melting point and in low concentration of NaCl the solution is in liquid phase. At higher concentrations, saturation leads to NaCl crystal or NaCl·2H$_2$O hydrates formation (shown on the right). Upon decreasing the temperature below the melting point, ice is formed pushing out the salt, in a process called brine rejection. When decreasing further the temperature below 251.9 K, ice and NaCl·2H$_2$O crystals coexist.

The temperature-concentration phase diagram of NaCl aqueous solution[29] is depicted in Fig. 1. Salt solutions are in liquid form at temperatures above the melting point and low concentration of NaCl, whereas increase of salt concentration beyond saturation initiates the formation of NaCl crystals. Depending on the temperature the resulting crystals can be either pure NaCl or NaCl·2H$_2$O hydrates (hydrohalites), shown in the right side of Fig.1. Upon decreasing temperature, ice formation occurs, leading to brine rejection, where the brine coexists with the ice crystals. The three phases meet at the *eutectic point* of NaCl[30] located at a concentration of 3.99 M (23.3% m/m) and a temperature of 251.9 K. Upon further cooling, the salt solution crystallizes and consists of a mixture of NaCl·2H$_2$O hydrates and ice crystals. The phase diagram of NaCl and that of other salt solutions has been investigated previously, and it is important for modeling sea and lake water[31], understanding the marine cloud microphysics[32] and improving cryobiological applications[33].

In this work, we investigate both experimentally and theoretically the crystallization of NaCl aqueous solutions and the brine rejection process using x-ray diffraction (XRD) at different temperatures and NaCl concentrations. A model is devised based on a linear superposition of saltwater, ice and hydrate crystals which allows us to reproduce the crystalline XRD pattern and quantify the amount of ice, brine and the formation of NaCl·2H$_2$O hydrates. The analysis is complemented by molecular dynamics (MD) simulations which allows us to capture evolution of ion concentration as a function of time and estimate the percentage of ions trapped inside the crystal lattice. Furthermore, the resulting structures are analyzed in terms of partial radial distribution functions for the different phases. Finally, we quantify the impact of ions in the ice structure by utilizing the tetrahedrality order parameter, as well as the formation of crystallites using the ion coordination parameter.

## 2. Materials and Methods

### 2.1. Experimental methods

A schematic of the experimental setup is depicted in Fig. 2A. The Bruker D8 VENTURE x-ray diffractometer was used to obtain diffraction data using a Molybdenum source (λ = 0.7107 Å). The temperature was regulated by a liquid nitrogen cryo-stream at the sample position and the cooling rate was set at 5 K/min. For each temperature the XRD pattern of the samples was measured sufficiently long as to ensure that equilibrium is reached. Microscope images were taken at different temperatures, as shown in Fig. 2B and C. These images were taken for a solution with 5% m/m NaCl concentration at temperatures 253 K and 233 K. One can observe the formation of polycrystalline ice domains and upon further cooling the appearance of salt microcrystals. The samples were measured in a quartz capillary (diameter 1 mm, wall thickness 10 μm). The NaCl aqueous solutions were prepared at different concentrations, expressed as the ratio of salt mass to total solution mass (m/m) in percent. The total sample volume of each solution prepared was 100 ml, whereas the volume illuminated by the beam (diameter 0.1mm) is estimated in the order of 1 μl. The NaCl solute employed was laboratory-grade reagent purchased from Sigma Aldrich and used without further purification in combination with Milli-Q water. An identical protocol was applied for all NaCl solutions measured at several temperatures and concentrations.

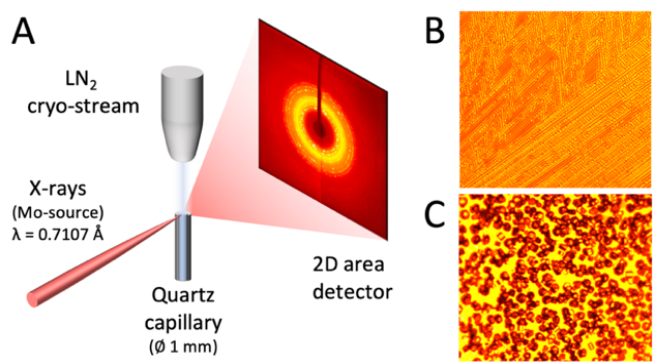

**Fig. 2** (A) Schematic of the experimental setup used and microscope images of the sample at two different temperatures at 5% m/m NaCl concentration, where the frame size is 1 mm (B) At 253 K is evident the formation of polycrystalline ice and (C) at 233 K we observe the appearance of salt microcrystals.

### 2.2. Molecular dynamics methods

We performed molecular dynamics (MD) simulations using the GROMACS package[34] (ver. 5.1.4), to investigate the exclusion of ions from a growing ice crystal. We used the OPLS-AA force field[35] for Na$^+$ and Cl$^-$ ions together with the TIP4P model of water[36]. The simulation was run at a temperature of T = 220 K with a total of $N_{H2O}$ = 13824 TIP4P molecules and $N_{Cl^-}$ = $N_{Na^+}$ = 100 Na$^+$ and Cl$^-$ ions, respectively. Ice growth was seeded with an ice block of $N_{ice}$ = 3456 TIP4P molecules corresponding to a crystal with dimensions of 5.40 nm × 4.65 nm × 4.39 nm. The concentration of the aqueous NaCl solution in contact with the ice block is thus 3 % m/m, i.e. 0.53 M (2.3 % m/m for the entire simulation box) initially chosen to be representative of the order of magnitude of typical seawater salt concentration. The melting temperature of the TIP4P model of water has previously been determined[37] to be $T_m$ = 232 ± 5 K and the simulation temperature thus corresponds to temperature difference from the melting point of pure TIP4P water of about ΔT = T - $T_m$ = -12 K at ambient pressure. We have not rescaled the concentration with respect to the saturation concentration of the model (see results section for an estimate), which can in fact deviate from the experimental values as indicated from previous detailed comparison between force fields[38,39].

The simulation was run in the NPT ensemble using the Bussi thermostat[40] and the Berendsen barostat[41] with semi-isotropic pressure coupling ($p_x = p_y \neq p_z$) to set the initial rectangular geometry of the simulation box to about (5.40 nm × 4.65 nm × 17.59 nm). Coulomb interactions were treated using the particle-mesh Ewald method, whereas real space Coulomb and van der Waals interactions were taken smoothly to zero between 0.87 and 0.9 nm. In addition, we used dispersion corrections and the simulation run for $t_{sim} = 1$ μs at a time step of $\Delta t = 2$ fs.

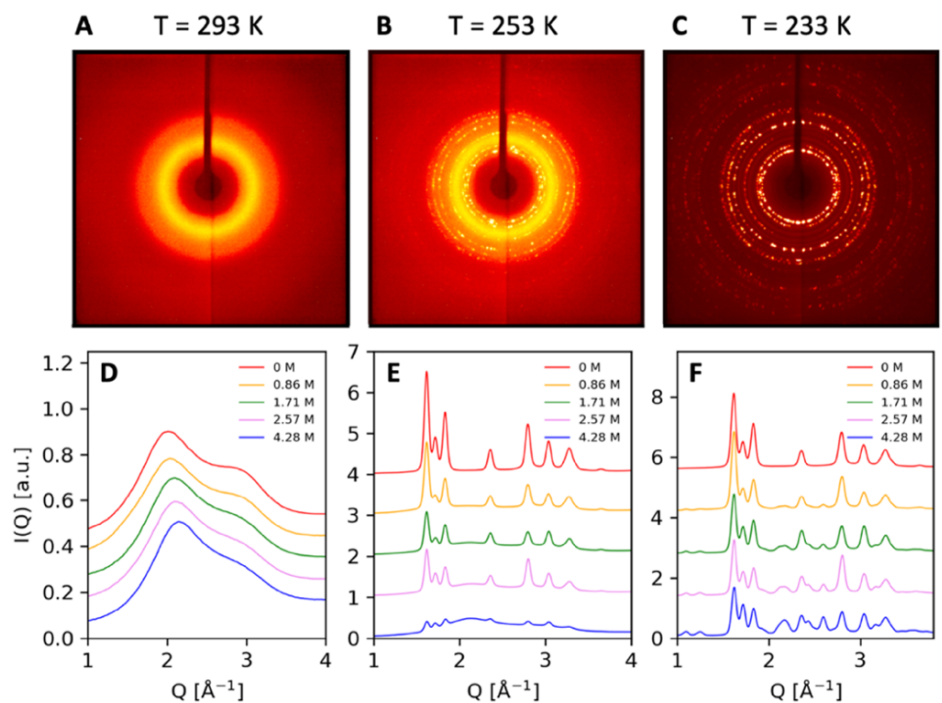

**Fig. 3** X-ray scattering patterns of NaCl solutions at 0.86 M (5% m/m) and temperatures (A) T = 293 K and (B) T = 253K and (C) T = 233 K. (D) Angularly integrated scattering intensity I(Q) of NaCl solutions at various concentrations indicated in the legend and temperature T = 293 K, (E) T = 253 K. (F) and T = 233 K.

## 3. Results

### 3.1. Experimental results

X-ray diffraction scattering patterns were obtained for a series of salt solutions, as well as for pure water and pure NaCl crystals. The XRD patterns recorded at temperatures T = 293 K, T = 253 K and T = 233 K and concentration 0.86 M (5% m/m) are shown in Fig. 3. At T = 293 K, the scattering pattern is uniform and broad, indicative of the sample being in liquid form, i.e. saltwater. At T = 233 K the appearance of sharp Bragg peaks indicates a polycrystalline structure arising from ice Ih and salt crystals. At the intermediate temperature T = 253 K, the scattering pattern exhibits coexistence of the liquid and crystalline forms manifested during the brine exclusion process. These three distinct structures observed correspond to different areas of the phase diagram presented in Fig. 1.

The angularly integrated scattering intensity I(Q) obtained for different salt concentrations, 0.86 M, 1.71 M, 2.57 M, 4.28 M (5% m/m, 10% m/m, 15% m/m, 25% m/m respectively) is shown in Fig. 3. Upon increasing concentration in the liquid at 293 K shown in Fig.3D, the first diffraction peak shifts and becomes less structured, in agreement with previous investigations[42,43]. This change in the scattering pattern is indicative of the influence of salt on the structure of the hydrogen bond network of water, seen also by x-ray absorption spectroscopy[44] and small-angle x-ray scattering[45]. The scattering intensity at T = 253 K shown in Fig.3E, corresponds to ice Ih coexisting with brine. In this condition, it is evident that the intensity of the peaks depends strongly on the concentration of salt and one can observe a decrease in the peak intensity with increasing salt concentration. This is due to the fact that by changing the amount of salt present in the solution, one effectively changes the relative ratio of ice and brine. For lower salt concentrations, the solution consists mainly of ice Ih and results in sharp Bragg peaks.

For T = 233 K the sample crystallizes completely and consists largely of a mixture of ice Ih and salt crystals, as can be seen in Fig. 3F for different salt concentrations.

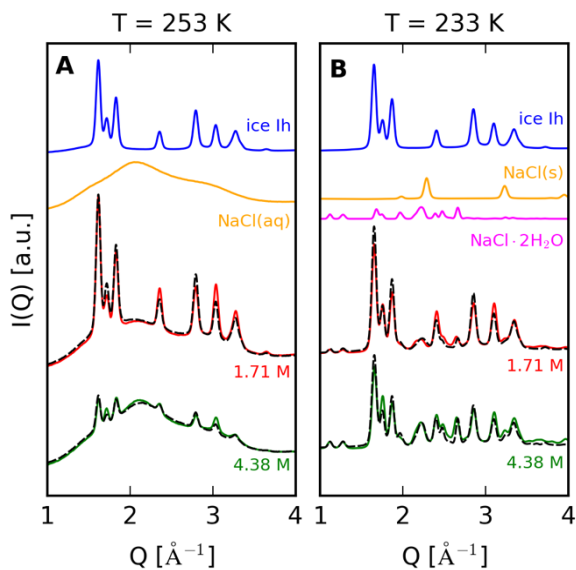
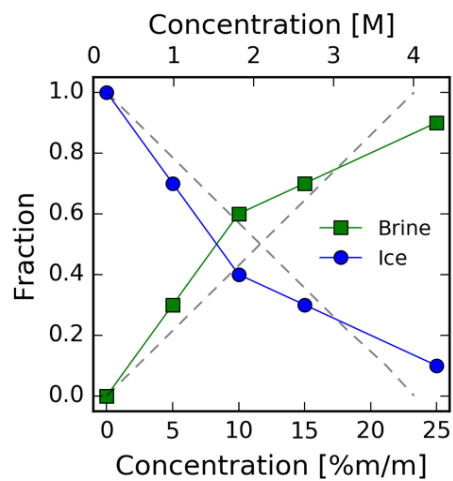

**Fig. 4.** (A) A model based on linear superposition of the different components is used to analyse the I(Q). The brine and ice coexistence at T = 253 K is modelled by combining ice Ih (blue solid line) with the NaCl aqueous solution at 4.28 M (orange solid line). The linear combination model (dashed line) is compared to the data at T = 253 K with concentration 1.71 M (10 % m/m) and 4.28 M (25 % m/m). (B) The same analysis for T = 233 K, where the components in this case are ice Ih (blue solid line), NaCl·2H$_2$O (magenta solid line) and NaCl crystals (orange solid line). The model (dashed line) is compared with the data recorded at concentrations 1.71 M (10 % m/m) and 4.28 M (25 % m/m).

Interestingly, upon increasing concentration, we observe the appearance of two new peaks at the lower momentum transfer region $Q$ = 1.11 Å$^{-1}$ and $Q$ = 1.278 Å$^{-1}$. The scattering pattern exhibits sharp Bragg peaks at this length scale, which are distinct from those present in the crystalline structure of ice Ih and the NaCl crystals and are a signature of the presence of NaCl·2H$_2$O hydrate formation.

In order to quantify the underlying contributions to the XRD diffraction pattern we construct a model comprising of a linear combination of the different components depending on temperature and concentration. Fig. 4A shows such an analysis for T = 253 K, where we observe coexistence of brine and ice Ih. The model utilizes two components shown in the upper row, where an offset has been introduced to facilitate the comparison. Here, the two components utilized are pure ice Ih measured at 250 K (blue solid line) and NaCl aqueous solution at T = 293 K with concentration 4.28 M (25% m/m) to model the saturated brine solution (orange solid line). The linear superposition of those two components indicated with a dashed line. When comparing this model with the data obtained at T = 253 K and concentrations 1.71 M (red solid line) and 4.28 M (green solid line) we observe good agreement with the experimental results.

**Fig. 5** The volume fraction of ice (blue squares) and brine (green circles) at T = 253 K, as a function of concentration indicated in both M (top axis) and mass percent m/m (lower axis). The dashed lines are the estimates from the phase diagram.

From the model described in Fig. 4, we estimate the fraction of ice and brine at each concentration, shown in Fig. 5. This fraction is calculated from the relative weights used in to reproduce the experimental scattering intensities The results from the linear superposition model are compared to the ice/brine fraction determined from the phase diagram (dashed lines) using the Lever rule[46,33]. More specifically, this rule states the volume fraction of two phases in equilibrium is proportional to the difference between the overall mass fraction of the two components at a given temperature. We observe that at concentrations 5 – 15 % m/m the model yields lower ice volume fraction that those obtained from the phase diagram, which can be attributed to the presence of ions in the ice lattice. Furthermore, one of the limitations of the model is that we use the NaCl aqueous solution at T = 293 K, whereas the data is taken at T = 253 K, which can introduce a shift in the first diffraction peak. An additional restriction of this approach is that this model does not include any interference (cross-terms) between the ice Ih and the brine, as the XRD pattern is the linear superposition of contributions arising from the two components.

A similar approach is used for the data at T = 233 K. The model consists of NaCl crystals (Fig.4 B – orange solid line), NaCl·2H$_2$O (Fig.4 B – magenta solid line) and ice Ih recorded at T = 233 K (Fig. 4 B – blue solid line). The dihydrate diffraction pattern is modeled as described in the Electronic Supplementary Information. In this case the model (dashed line) again exhibits quantitative agreement with the measured XRD pattern and indicates that the scattering intensity is largely due to the formation of NaCl dihydrates, which is consistent with previous observations[32].

## 3.2. Simulation results

Snapshots of the MD at different times of the trajectory at T = 220 K are shown in Fig. 6. The Na$^+$ ions are denoted by dark blue spheres and Cl$^-$ ions by turquoise spheres. Initially, the simulation was seeded with a block of pure ice Ih with the basal plane exposed to the NaCl liquid solution (t = 0). Upon crystallization the majority of the ions is expelled into the liquid form. However, there is a small fraction of ions trapped in the ice lattice, consistent with previous MD simulations[26] where it was seen that more Cl$^-$ is incorporated in the lattice than Na$^+$.

The concentration of ions in ice and in the brine is shown as a function of time in Fig. 7 (see also Electronic Supplementary Information). The rise in concentration coincides with the potential energy decrease due to freezing, shown in gray (right-hand side y-axis). The potential energy of the system (Fig. 7) decreases rapidly upon ice growth and the decrease at about 200 ns, when we observe the formation of small salt crystallites in the remaining brine liquid (see cluster analysis below). The minor increase of ion concentration in ice to about 0.5% m/m is due to ion inclusion in the ice lattice.

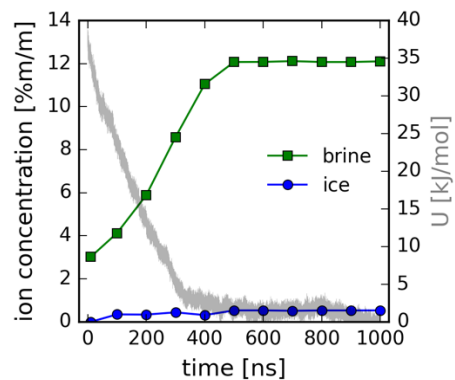

**Fig. 7** Ion concentration (left y-axis) in ice (blue circles) and brine (green squares) and the potential energy in the MD simulation (gray line – right y-axis) as functions of time

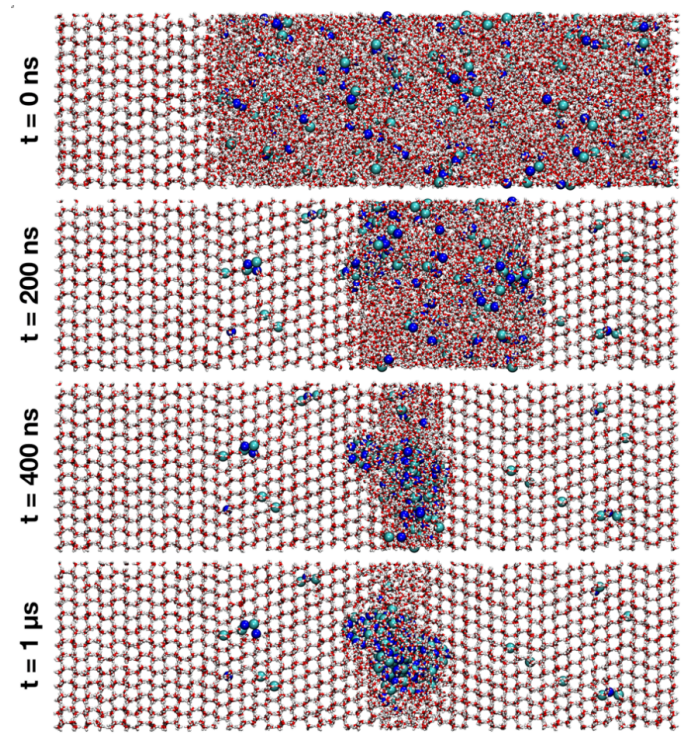

**Fig. 6** Snapshots obtained from MD simulations at T = 220 K showing ice crystal growth at times t = 0, 200 ns, 400 ns and 1 μs. A small fraction of ions get trapped within the ice structure while the majority of ions are expelled into the solution phase of increasing salt concentration. Na$^+$ ions are represented by dark blue, while Cl$^-$ ions by turquoise spheres. The snapshots were rendered with VMD[47].

This value is significantly reduced from the initial concentration 3% m/m, and near the standard salinity limit of fresh water, typically 0.1% m/m for drinking water and 0.5% m/m for agricultural use. We note here that the simulation using OPLS-AA parameters for the ions together with the TIP4P water potential underestimates the experimental solubility and saturation concentration, which is 6.14 mol/kg (0.599 M) at T = 273 K[48]. From the concentration at the time of the onset of heterogeneous crystallite formation around 200 ns, we estimate the saturation concentration for the OPLS-AA/TIP4P combination to be roughly 1 M at T = 220 K. This is consistent with similar force field combinations, e.g. AMBER-99 or OPLS-AA ion parameters with TIP3P water, yield also solubility values between 1 - 2 mol/kg[39] (corresponding to roughly 1 - 2 M). Furthermore, this issue has been potentially improved in relatively recent force field parameters for NaCl ions[49], as salt crystal formation was not observed in the brine phase at similar conditions[26].

In order to investigate the formation and local structure of the NaCl dihydrate crystals encapsulated in ice Ih, we calculate the radial distribution function $g(r)$ shown in Fig. 8. The panel (A) shows the total $g(r)$ as a function of simulation time, where the changes observed are mainly due to freezing. In Fig.8B is shown the O-O contribution comparing the various components, such as the crystalline (s) and the aqueous environment (aq). A distinction is made between the self-correlations of the different atoms, such as the O-O, Cl-Cl and Na-Na, with the corresponding cross-correlations, such as Na-Cl, O-Cl and O-Na. The O-O radial distribution function $g(r)$ of the crystal (solid line) resembles that of ice Ih, with the addition of disorder due to the water-brine interaction. The corresponding brine at the beginning of the trajectory is liquid phase is depicted as the dashed line. Both the Cl-Cl and Na-Na correlations exhibit a peak at $r$ = 4 Å which is consistent with the atomic spacing in NaCl crystals where the FCC lattice spacing is d = 5.64 Å, and the shortest Cl-Cl and Na-Na distance is $d/\sqrt{2} \approx 4$Å. The Na-Cl contains a dominating peak at $r$ = 2.8 Å and in addition two peaks at $r$ = 4.9 Å and 6.35 Å, which correspond to Na-Cl spacing in the salt crystal lattice. This is an indication of near ion-ion ordering that resembles locally the crystal.

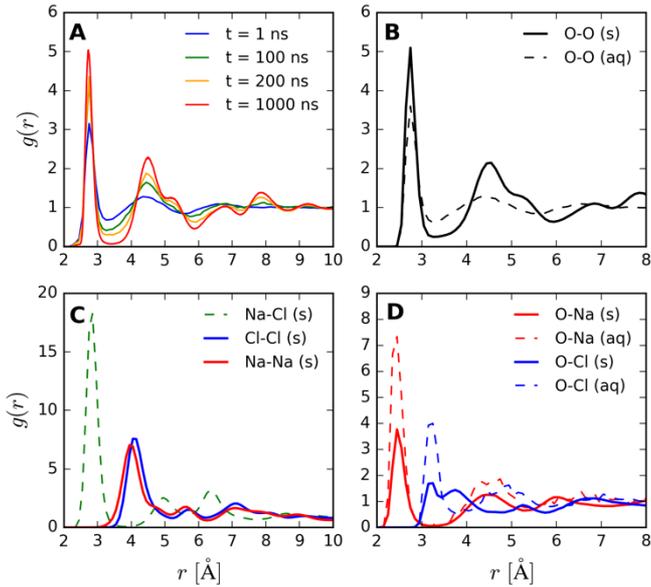

**Fig. 8** The radial distribution function $g(r)$ obtained from the MD simulations. (A) The time evolution of the total $g(r)$ (B) the O-O of the crystalline (s) and aqueous (aq) components and (C) Cl-Cl, Na-Na as well as the cross-correlations between Na-Cl and (D) O-Cl, O-Na for both crystalline (s) and aqueous (aq) components.

The ion-oxygen cross-correlations in the radial distribution function $g(r)$ (Fig. 8D) contains information about the O-Cl and O-Na components in the crystal (s), which partially resembles the radial distribution of ions in liquid NaCl aqueous solutions (aq) [50,51]. These contributions arise from an ion (Na$^+$ or Cl$^-$) surrounded by water molecules, occurring within the hydrates and at the boundaries of the NaCl crystallite with water. The first peak reveals the first neighboring water molecules around the ion, which in the case of Na-O correlation corresponds to distances of $r$ = 2.45 Å and for Cl-O of $r$ =3.18 Å. This difference is due to the influence of the positive charge of the Na$^+$ which attracts the oxygen, whereas in the case of Cl$^-$ negative charge attract the hydrogen and results in larger O-Cl distances. Interestingly the O-Cl correlation features an additional peak around 3.75 Å which is attributed the hydrate formation (see Electronic Supplementary Information for the $g(r)$ of dihydrate).

We quantify the effect of ions on the local ice structure by calculating the tetrahedrality parameter[52], which enables us to quantify the orientational order of the water molecules in the simulation as function of distance to the closest ion. The tetrahedrality parameter is defined as

$$q_i = 1 - \frac{3}{8}\sum_{j=1}^{3}\sum_{k=j+1}^{4}\left(\cos(\psi_{jik}) + \frac{1}{3}\right)^2,$$

where $\psi_{jik}$ is the angle formed by the lines joining the oxygen atom of a given molecule and those of its nearest neighbors $j$ and $k$ ($\leq 4$). This analysis, shown in Fig. 9A, reveals a high average tetrahedrality parameter of about $\langle q_i \rangle \approx 0.94$ for molecules that are part of the ice, whereas we see a difference between the Cl-O and Na-O, consistent with the corresponding radial cross-correlations in the radial distribution function. Furthermore, from the time evolution of the tetrahedral parameter (Fig.9B) we note that the major changes occur in the range $0.7 < \langle q_i \rangle < 0.9$ for O–ion distances in the interval 4–15 Å and at early simulation times. In this intermediate distance range interval, $\langle q_i \rangle$ increases over time indicating the molecules less affected by ions become part of the ice phase.

Furthermore, in order to quantify the formation of the crystallites and understand the underlying mechanism, we define a simple "ion coordination parameter" $n_l = \frac{N_l}{N_{max}}$ as the number of neighbor ions $N_l$ within a cut-off distance of $d_c = 3.4$ Å and divide by the maximum number $N_{max}$ of counter-ions within this distance for the ideal ionic crystal of cubic symmetry, i.e. $N_{max} = 6$. The parameter defined above does not directly probe the local symmetry; yet, due to strong Coulomb interactions, this simple approach can be justified. The so-defined parameter takes average values close to unity for a NaCl crystal and average values close to zero for dissociated ions in solution where only temporary ion-pairing is expected to make this number slightly larger than zero. Intermediate values of $\langle n_l \rangle$ indicate that there are ions in contact with their counter-ions yet there is no complete coordination as compared to coordination in cubic symmetry. For a NaCl·2H$_2$O crystal, the coordination parameter is $n_l = 0.586$, whereas in the case of the simulation we observe that $n_l$ levels off at $n_l = 0.467$ after 400 ns approaching the hydrate values. Any deviation here is attributed to ions in the surface of the crystal, which will have smaller ion coordination numbers.

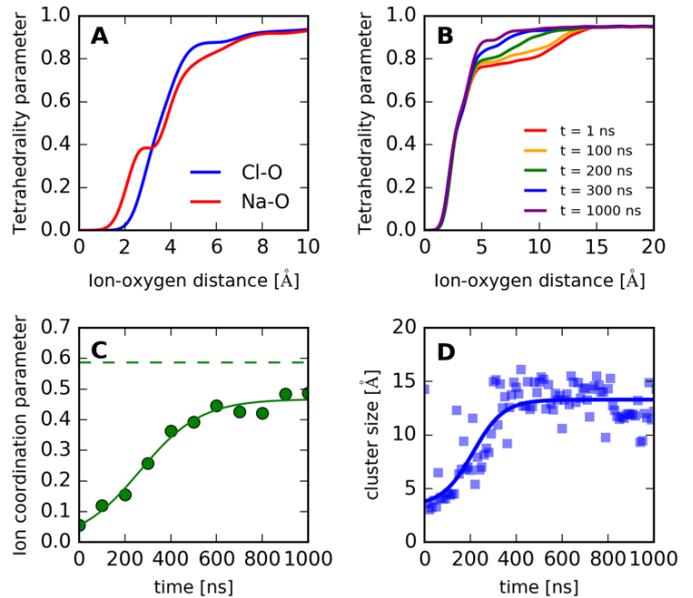

**Fig. 9** (A) Tetrahedrality parameter as function of distance between the water's oxygen and the closest ion, for Cl-O and Na-O. (B) The tetrahedrality parameter for different times, indicated in the legend. (C) Ion coordination parameter as a function of time (green circles) and the same parameter calculated for a NaCl·2H$_2$O crystal (dashed line). (D) The crystallite size quantified by cluster analysis as a function of simulation time.

In addition, we calculate the crystal size by performing cluster analysis[53,54] (see Electronic Supplementary Information). The cluster size as a function of time is shown Fig. 9D, which reaches a constant value on a similar time scale as the ion coordination parameter, indicating the formation of ion clusters with average size of roughly 13 Å.

## 4. Conclusions

In summary, we have investigated the crystallization of NaCl aqueous solutions using XRD for different temperatures (293 K, 253 K, 233 K) and NaCl concentrations (0.86 M, 1.71 M, 2.57 M, 4.28 M). The brine rejection mechanism is being discussed and we observe that a model consisting of a linear combination of the x-ray diffraction intensity of ice Ih crystals and NaCl aqueous solution reproduces well the experimental data and allows us to quantify the fraction of recovered ice and brine at each condition, which is indicative of the freeze desalination efficiency. In addition, the regime where ice Ih coexists with NaCl crystals is analyzed. In this case a model consisting of two components, ice Ih and NaCl dihydrate crystals, reproduces well the observed diffraction pattern and helps us identify the signature of hydrate formation in the scattering intensity.

In order to model the salt-exclusion and formation of NaCl dihydrates, we performed MD simulations of the NaCl aqueous solution crystallization process. The simulation results were analyzed in terms of radial distribution functions for oxygen-oxygen, oxygen-ion and ion-ion correlations. We observe signatures of hydrate formation in the O-Cl correlation. Furthermore, we quantify the impact of ions on the ice structure by calculating the tetrahedrality parameter as a function of ion-oxygen distance. The structure of the resulting crystallite is analyzed using the ion coordination parameter and cluster analysis, which indicate that hydrate formation takes place within 400 ns. From the simulations we can estimate the concentration of ions trapped inside the ice, which is approximately 0.5% m/m. This value is reduced significantly from the initial NaCl concentration 3 % m/m and is on the fresh water salinity limit, which is typically in the order of 0.1% m/m for drinking water and 0.5% m/m for agriculture.

The present investigation provides a valuable insight into the limitations of freeze desalination by employing an x-ray-based detection method of the ions trapped inside ice in the form of crystal impurities. This study sets the stage for future investigations using ultrafast x-ray scattering at x-ray free-electron lasers that would allow to follow the brine rejection in-situ with nanosecond resolution. It will also be interesting to resolve spatially the formation of hydrates by utilizing the new diffraction-limited synchrotron sources, where a nano-focus x-ray beam would allow to quantify experimentally the size of the salt crystallites.

## Acknowledgements

We acknowledge financial support from the Swedish Research Council (VR starting grant under project no. 2019-05542) and from the European Research Council (ERC advanced grant WATER under project no. 667205). This research is part of the MaxWater initiative of the Max-Planck Society. We would like to thank L. G. M. Pettersson, T. J. Lane and K. Amann-Winkel for useful comments on the manuscript. I.T. was additionally supported by the Erasmus+ program of the European Union. The simulations were performed on resources provided by the Swedish National Infrastructure for Computing (SNIC) at the National Supercomputer Centre (NSC) and the High-Performance Computing Centre North (HPC2N).

# Brine rejection and hydrate formation upon freezing of NaCl aqueous solutions

Ifigeneia Tsironi,[a][†] Daniel Schlesinger,[b][†] Alexander Späh[a], Lars Eriksson[c], Mo Segad[c] and Fivos Perakis*[a]

[†] *These authors contributed equally to this work.*
[a] *Department of Physics, AlbaNova University Center, Stockholm University, 114 19 Stockholm, Sweden*
[b] *Department of Environmental Science & Bolin Centre for Climate Research, Stockholm University, 114 18 Stockholm, Sweden*
[c] *Department of Materials and Environmental Chemistry, Stockholm University, 106 91 Stockholm, Sweden*
* Corresponding author: f.perakis@fysik.su.se

## 1. Concentration estimation

We have estimated the mass percent concentration from the MD simulations by counting the number of water molecules $n_{H_2O}$ and ions $n_{Na}$, $n_{Cl}$ within a given volume and using the corresponding molecular/atomic masses $m_{Na}$, $m_{Cl}$ and $m_{H_2O}$:

$$c[m/m\%] = \frac{n_{Na} \cdot m_{Na} + n_{Cl} \cdot m_{Cl}}{n_{H_2O} \cdot m_{H_2O} + n_{Na} \cdot m_{Na} + n_{Cl} \cdot m_{Cl}}$$

Each frame was analysed separately and the volume of the brine was selected by taking vertical planes perpendicular to the ice growth axes, shown in Fig. S1. The straight boundaries were chosen due to simplicity, which seem to match pretty well the ice growth front.

In addition, we estimate the molar concentration, in the following way:

$$c[mol/L] = \frac{n_{Na} + n_{Cl}}{2 \cdot N_A \cdot V}$$

where $N_A$ is the Avogadro number and V the volume. The molar concentration as a function of time is plotted in Fig. S2.

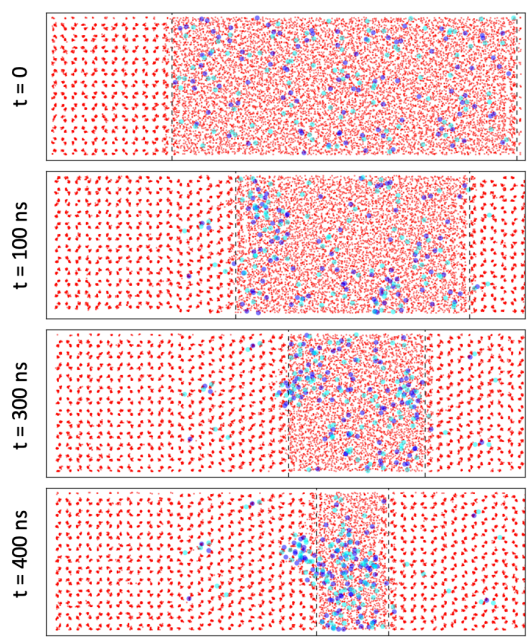

**Fig. S1.** Detail on the calculation of the concentration, where the dashed lines indicate values where the volume estimation.

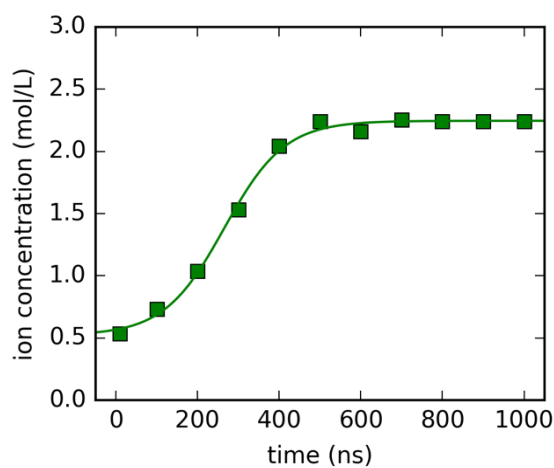

**Fig. S2.** The ion molar concentration of brine as a function of simulation time.

## 2. Cluster analysis

The cluster analysis was performed in the following way: we define a cluster consisting of ions that are linked with their first neighbor counterions, according to the ion coordination parameter. We kept only clusters consisting of more than 2 ions and in addition we merge neighbouring clusters, if their centers are separated by distances smaller than 10 Å. The last step assumes hierarchical clustering and the cut-off was selected by performing the analysis on a NaCl·2H$_2$O crystal, as detailed below. This step is necessary for identifying hydrate clusters due to the present of water molecules as part of the crystal.

The cluster analysis was tested the crystal dihydrate structure, shown in Fig.S3. The upper left panel shows the actual dihydrate structure. The cut-off values are indicated at the corresponding panels. The identified ion clusters are highlighted in different colours. One can see that upon increasing the cut-off parameter, eventually the whole crystal is identified as one cluster.

After identifying the appropriate values for the cut-off distance the clustering algorithm was applied on the MD trajectory. For each frame only the size and number of the largest cluster was identified. The results are shown in Fig. S4, where different frames are displayed with the identified ion clusters highlighted in colour. At early times only minor clusters are found (50 ns). As crystallization progresses and the brine concentration increases the average cluster size in progressively increases and finally peaks after 400 ns. We also show the number of molecules per cluster in Fig. S5, which follows the same trend with the cluster size shown in the main text.

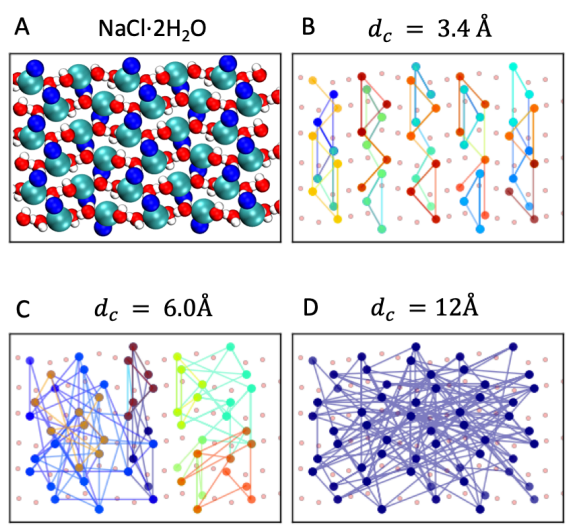

**Fig. S3.** Test of the clustering algorithm on a NaCl·2H$_2$O crystal. (A) the crystal, where the Na$^+$ ions are shown in blue, the Cl$^-$ ions in cyan and the oxygens and hydrogens are shown in red and white. (B) The result from the cluster analysis applied on the crystals with a cut-off distance of 3.4Å. The different identified ion clusters are labelled with different colours. (C) and (D) The clustering analysis applied with cut-off distances 6.0 Å and 12 Å.

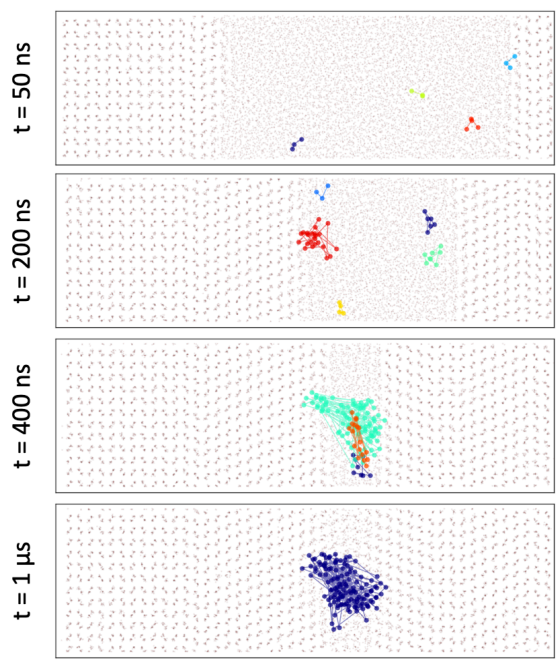

**Fig. S4.** Different frames from the cluster analysis applied on the MD trajectory. The simulation times are indicated in the right hand side and the corresponding identified ion clusters are highlighted in different colors.

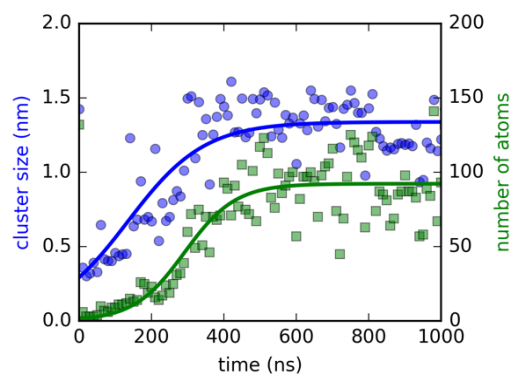

**Fig. S5** The cluster size and the number of atoms per cluster as a function of time.

## 3. Hydrate radial distribution function

A NaCl·2H$_2$O crystal was optimized with TIP4P and the radial distribution function was calculated in order to compare with the crystallite observed in the main text. The results are shown in Fig. S6. The peak at the O-Cl component (marked with a *) is also present in the MD simulation results presented in the main paper, which we consider a signature of the hydrate formation. This peak is not present in the corresponding liquid state or in the NaCl crystal.

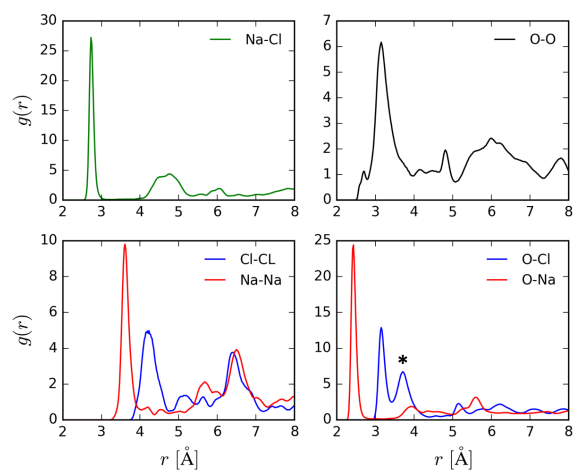

**Fig. S6** The radial distribution function $g(r)$ obtained from a NaCl·2H$_2$O crystal, which was optimized with TIP4P. The different panels correspond the different components, indicated in the legend. The (*) indicated on the O-Cl component is also indentified in the $g(r)$ presented in the main text, which is attributed to hydrate formation and is not present in the brine.

## 4. Hydrate scattering intensity

The diffraction pattern of NaCl dihydrate (Fig. S7) was modeled based on the refined structure determined by x-ray diffraction[1] (see inset Fig. S7). A Gaussian line shape with (FWHM) 0.05 Å$^{-1}$ was used to model the contributions to the diffraction peaks from a convolution of the instrumental line broadening with the crystal size. In addition, the Debye-Waller factor DWF = $\exp(-BQ^2)$ was used in order to attenuate the peaks intensity due to thermal motion, where B is the fitting parameter for each temperature and concentration. The obtained diffraction pattern with (blue solid line) and without (black dashed line) the DWF (red solid line) is shown in Fig. S7 for a typical value of B = 0.4 Å$^2$.

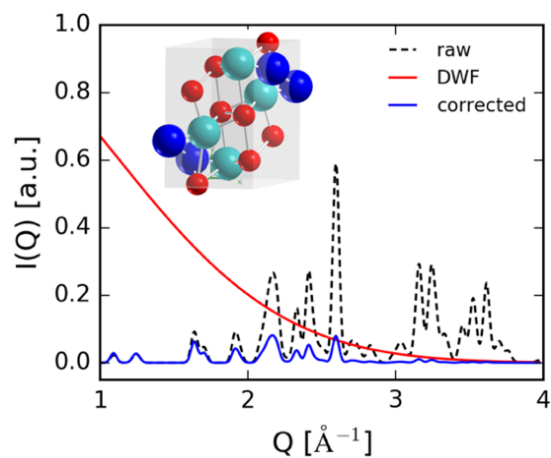

**Fig. S7** The modelled XRD pattern of NaCl dihydrate obtained by structural refinement [1]. The Debye-Waller factor (DWF) is included to model the intensity attenuation due to thermal motion. The data obtained with (blue sold line) and without (black dashed line) the DWF are shown here. The inset depicts the structure of the NaCl·2H$_2$O hydrate, where O atoms are represented by red spheres, Na$^+$ ions by dark blue, Cl$^-$ ions by turquoise spheres and H atoms are not shown.